%% file: main.tex
\documentclass[conference]{IEEEtran}
\IEEEoverridecommandlockouts
\usepackage{cite}
\usepackage{amssymb,amsfonts}
\usepackage{graphicx}
\usepackage{pgfplots}
\usepackage{amsmath}
\pgfplotsset{compat=1.15}
\usepackage{xurl}
\usepackage{tikz}
\usetikzlibrary{arrows.meta, positioning, shapes.geometric, patterns}
\usepackage{listings}
\usepackage{subfig}
\def\BibTeX{{\rm B\kern-.05em{\sc i\kern-.025em b}\kern-.08em
    T\kern-.1667em\lower.7ex\hbox{E}\kern-.125emX}}

\begin{document}

\title{ALARM: Adaptive Layer-Aware Resource Management for Power-Efficient vRANs}

\author{
\IEEEauthorblockN{
Ali Srour,
Farzad Veisi,
Sami Taktak,
Vania Conan
}
\IEEEauthorblockA{
CEDRIC Lab, CNAM, Paris, France \\
\{ali.srour, farzad.veisi-goshtab, sami.taktak, vania.conan\}@cnam.fr
}
}

\maketitle

%========================================================
% Section : Abstract
%========================================================

\begin{abstract}
The transition to virtualized Radio Access Networks (vRAN) enables dynamic power control through fine-grained CPU resource management. However, existing approaches treat gNB as a monolithic entity, failing to exploit the heterogeneous computational characteristics of different protocol layers. This paper proposes ALARM, a layer-aware adaptive resource management framework for constrained 5G vRAN. ALARM decomposes the gNB into functional layers representing distinct RAN tasks with heterogeneous computational demands, and controls CPU resources at the layer level based on processing weights. The framework monitors per-layer performance to detect violations and responds by scaling only the affected layer. This targeted adaptation avoids the over-provisioning inherent in traditional uniform scaling approaches. Experimental validation on two constrained platforms demonstrates up to 33\% power reduction versus non-optimized baseline and 19\% beyond uniform approaches, with dynamic adaptation achieving 9.8\% additional savings while preserving strict real-time guarantees.

\end{abstract}

%======================================================
% Section I : Introduction
%======================================================

\section{Introduction}\label{section:introduction}

The evolution toward vRAN is motivated by the need for greater flexibility and cost-efficient deployment in modern cellular networks. However, this architectural shift also changes the power consumption profile of base stations. In traditional hardware-based RANs, the Radio Unit (RU) typically dominates the energy budget, accounting for 60-80\% of total power consumption \cite{Powerofradio}. In contrast, vRAN introduces additional software processing overhead, increasing the contribution of baseband processing to the overall energy consumption \cite{foukas_concordia_2021}. This effect is particularly pronounced in small-cell deployments, where lower transmit power reduces radio consumption while computational demands remain largely unchanged. As a result, processing power can become comparable to, or even exceed radio power consumption ~\cite{ge_energy_2017}, making processing efficiency a key lever for reducing the overall energy consumption of vRAN systems, especially in constrained cells.
%------Background and Motivation added

A 5G gNB comprises multiple protocol layers with fundamentally different computational characteristics. The physical layer dominates processing time~\cite{wei_5gperf_2022,foukas_concordia_2021}, with load depending strongly on radio parameters and allocated Physical Resource Blocks (PRBs)~\cite{tran_understanding_2017}. This imbalance reveals significant energy-saving potential. Traditional uniform resource control approaches satisfy peak demands of compute-intensive layers but over-allocate to lighter layers. Instead, resource provisioning must be performed at a layer granularity by determining the requirements of each protocol layer. To exploit this opportunity, we examine how vRAN software executes and is scheduled on general-purpose CPUs.

%---vRAN Execution: 
In modern 5G deployments, gNB baseband functions are realized as software processing pipelines, where RAN tasks (e.g., channel coding, MAC scheduling) execute as queue-based user-space threads managed by a worker model \cite{foukas_concordia_2021}. This design is common in vRAN implementations (e.g., Intel FlexRAN~\cite{li20185gl2}, OpenAirInterface, srsRAN \cite{srsRAN}). RAN threads are then managed by general-purpose OS schedulers (e.g., CFS, FIFO), which do not optimize for power efficiency. Beyond power efficiency, these threads face stringent timing requirements. Physical layer threads execute signal-processing tasks that must be completed within a Transmission Time Interval (TTI), defined by 3GPP slot structure~\cite{3GPP38211}. The consequences of violating deadlines range from performance degradation (e.g., dropped throughput), to system failure (e.g., vRAN software crashing).
%---Motivation and solution:

These real-time requirements constrain the applicability of conventional CPU power-management mechanisms in vRAN systems. Standard approaches have proven effective for real-time applications, however their impact on vRAN is constrained by OS scheduling policies that lack RAN awareness ~\cite{foukas_concordia_2021}. Prior research on vRAN power management spans adaptive frequency control, core affinity optimization, and traffic-driven resource provisioning ~\cite{kalia_towards_nodate, huang_reinforcement_2024, crespo_energy-aware_2025, urumkar_power_2025, urumkar_improving_2023, leonelli_dynamic_2025, lozano_airic_2024}, demonstrating measurable energy savings. However, applying identical control levels across all protocol layers, regardless of their vastly different computational intensities creates an inherent efficiency ceiling. Our work extends these proven management techniques with a new dimension, a layer-aware traffic-driven resource management framework, termed ALARM. ALARM leverages the gNB structuring into functional processing layers, each characterized by a weight representing its target CPU share. Resource orchestration policies are assigned to each layer. Dynamic adaptation incorporates violation identification and applies corrective configurations exclusively to the affected layer. 

We evaluate ALARM using resource-constrained platforms representative of small-cell deployments. ALARM achieves up to 33\% power reduction by eliminating over-provisioning, while maintaining the strict real-time guarantees across varying traffic loads and achieving 9.8\% additional savings over uniform adaptation scaling approaches. 
The main contributions of this paper are:
\begin{itemize}
\item We introduce a fine-grained, power-efficient resource management approach that leverages the layered design of vRANs and treats protocol layers independently through processing weights, thereby replacing monolithic orchestration with proportional resource control.

\item We design a per-layer adaptation framework that detects layer-level violations using violated data volume and dynamically adjusts resources based on real-time workload, scaling only affected layers.

\item We validate the approach on real 5G vRAN testbeds using resource-constrained platforms, demonstrating significant power reduction.
\end{itemize}

%========================================================
% Section II : Related Work
%========================================================

\section{Related Work}
%---Standard Linux solutions
Standard CPU power management relies on two primary mechanisms: dynamic frequency scaling (P-states) and sleep states (C-states). These mechanisms typically operate at control timescales of 10+ ms~\cite{wu2015energy}, which are orders of magnitude slower than vRAN's sub-millisecond slot deadlines. Consequently, vRAN deployments~\cite{li20185gl2} carefully constrain or entirely disable Linux-based power management features to avoid performance degradation. While enabling these techniques effectively reduce power consumption in general-purpose workloads, their effectiveness in vRAN depends critically on workload accuracy. OS schedulers, lacking RAN-specific awareness, introduce overhead through context switching, thread migration, and cache misses~\cite{foukas_concordia_2021}. This scheduling inefficiency triggers unnecessary activity in uncore components (L3 cache, memory controller, PCIe), forcing Dynamic voltage and  frequency scaling (DVFS) to elevate frequencies system-wide and limiting energy savings. This motivates RAN-aware power management with real-time guarantees.

%----------Related Research Work and Limitations:
Prior research on base station energy efficiency has primarily focused on hardware-based systems using sleep modes~\cite{wu2015energy}, which are not directly applicable to vRAN's general-purpose CPU architectures. Recent vRAN power management research has explored multiple directions, each with distinct limitations.

Kalia et al.~\cite{kalia_towards_nodate} identify the high response latency of standard P/C-states (+60 $\mu$s) and propose fast kernel-based reactive frequency scaling for vDU. Building on this direction, Huang et al.~\cite{huang_reinforcement_2024} use reinforcement learning to adjust CPU frequency based on Packet Data Convergence Protocol (PDCP) traffic load and block error rate (BLER). While these reactive approaches demonstrate the feasibility of dynamic frequency control, they face fundamental constraints in metric selection and resource resolution. PDCP operates at timescales that miss sub-millisecond traffic bursts, while BLER reflects radio channel conditions rather than actual computational load. More critically, these approaches apply uniform frequency scaling across all cores despite uneven load distribution, which could lead to suboptimal efficiency.

Another research direction focuses on mapping gNB functions to physical cores to improve resource utilization. Crespo et al.~\cite{crespo_energy-aware_2025} leverage O-RAN dApps for core affinity and frequency scaling, while Urumkar et al.~\cite{urumkar_power_2025, urumkar_improving_2023} propose CPU sleep modes for entire gNB stacks. While these techniques demonstrate measurable improvements in core utilization, they operate at a coarser granularity than what vRAN's architecture permits. Specifically, these approaches lack layer-specific allocation, explicit violation detection, and traffic-driven adaptation. Additionally, sleep mode approaches are difficult to apply to RAN and require low-latency mechanisms to avoid crashes~\cite{kalia_towards_nodate}. Recent work has also explored dynamic resource allocation at the DU/CU component level, with Leonelli et al.~\cite{leonelli_dynamic_2025} and Lozano et al.~\cite{lozano_airic_2024} proposing traffic-based resource provisioning between functional components. However, component-level granularity remains too coarse to exploit the fine-grained heterogeneity within the baseband processing pipeline itself.

%========================================================
% Section III : Solution proposed
%========================================================

\section{ALARM: Adaptive Layer-Aware Resource Management Framework} \label{section:solution}

ALARM builds on established power management techniques and extends their scope by introducing layer-level energy efficiency as a complementary dimension. Rather than replacing existing mechanisms, our approach enhances them by monitoring actual computational load through direct per-layer violation detection and adapting resources independently for each affected layer. This strategy preserves the real-time guarantees that prior work has shown to be critical, while exploiting the computational heterogeneity an opportunity that becomes accessible through vRAN's software architecture.

\subsection{Layer-Aware gNB Processing Model}\label{section:layer_aware_gnb}

To treat all layers differently, we model the gNB as a collection of heterogeneous processing layers. We consider a gNB composed of a finite set of processing layers $\mathcal{L} = \{L, D, U, Up\}$, corresponding to low-PHY (L) handling lower physical layer processing, DL high-PHY (D) and 
UL high-PHY (U) managing uplink and downlink higher physical layer tasks respectively, and upper layers (Up) executing layer 2/3 protocol functions, where each layer $\ell \in \mathcal{L}$ is implemented as one or more threads. While we use this four-layer decomposition for clarity, the modeling framework is adaptable and can accommodate more granular layer definitions based on RAN software architectures.

We formalize the layer-aware resource management as follows. For each layer $\ell \in \mathcal{L}$, we define the number of threads $N_\ell \in \mathbb{N}^+$, the CPU frequency $F_\ell \in [F_{\min}, F_{\max}]$ assigned to cores hosting layer $\ell$, and the set of CPU cores $C_\ell$ to which layer $\ell$ is mapped. The resource control for the entire gNB is then represented by the tuple $\mathcal{R_\ell} = \{(N_\ell, F_\ell, C_\ell) : \ell \in \mathcal{L}\}$.

Each layer maintains an approximately constant share of total CPU utilization across traffic types \cite{wei_5gperf_2022}. This consistent ratio enables defining per-layer processing weights: each layer $\ell$ is associated with weight $w_\ell$ representing its typical CPU demand share, normalized such that $\sum_{\ell \in \mathcal{L}} w_\ell = 1$, enabling proportional resource control based on actual computational demands. The processing weights $w_\ell$ guide initial resource provisioning and core-affinity decisions $C_\ell$, while actual resource allocation $\mathcal{R_\ell}$ adapts dynamically based on traffic patterns, violations, and power optimization.

\subsection{Framework  Design}\label{section:framework_design}

\begin{figure}[ht]
  \centering
  \includegraphics[width=0.8\linewidth]{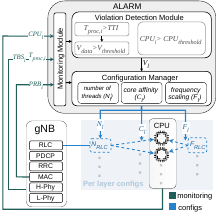}
  \caption{ ALARM's design overview illustrating monitoring of RAN/system metrics, and applying per layer configurations}
  \label{fig:framework}
  %\vspace{-.9em}
\end{figure}

This part illustrates the framework architecture and component interactions shown in Fig.~\ref{fig:framework}. The framework consists of three main components: monitoring module, configuration manager module, and violation detection module.% 

First, the monitoring module interfaces with the gNB software stack through two mechanisms, internal instrumentation where slot-level metrics are exposed by vRAN software through the open interfaces including $PRBs$, transport block size ($TBS$) and physical processing time ($T_{\text{proc}}$), and external monitoring with system-level metrics (CPU utilization per thread, frequency per core). These measurements are provided to other modules to have visibility on workload variations, deadline violations and per-layer processing bottlenecks.  

Second, a configuration manager module decides which configuration to apply and enforces action decisions for each layer. Based on layer monitored state the module adjusts the number of worker threads ($N_\ell$), CPU core affinity ($C_\ell$), and operating frequency ($F_\ell$) independently per layer. For example, a thread allocation configuration across all layers is represented as ($N_{\mathrm{L}}$, $N_{\mathrm{U}}$, $N_{\mathrm{D}}$, $N_{\mathrm{Up}}$).

Third, violation detection module analyzes monitored metrics to detect physical layer deadline miss, and traffic load. The control loop is executed at a TTI timescale compatible with vRAN dynamics, allowing the framework to react to sustained workload changes while avoiding instability due to short-term fluctuations. For a better understanding, the component detailed behaviors are presented in subsequent subsections.

\subsection{Traffic Aware and Per-Layer Violation Detection}\label{section:traffic_aware}
 
A critical challenge in dynamic vRAN resource management is determining when to scale resources up or down. Aggressive resource reduction saves energy but risks violating real-time deadlines. Given the complexity of processing physical tasks within sub-millisecond slots, per-TTI computational load is high and primarily driven by the number of allocated PRBs~\cite{GarikipatiCoNEXT16}. When processing capacity is insufficient, whether due to limited capacity of processors or insufficient number of executors, packets in the affected slot are dropped~\cite{foukas_concordia_2021}. This makes violation detection essential for maintaining real-time guarantees while optimizing power consumption. The violation detection module addresses this challenge through per-layer violation-based feedback. The module gets $T_{\text{proc}}$ from the monitoring module and compares it against the TTI budget to identify individual slot violations: $V_{\text{slot},i} = 1$ if processing time exceeds the TTI deadline, and $0$ otherwise, where i is slot index. However, not all deadline violations require resource reallocation. Individual violations often stem from transient factors like scheduling jitter or cache misses. Reacting to every violation causes control instability and energy waste. To distinguish transient violations from genuine capacity shortage, the framework computes violated data volume rather than counting violations. This approach weights violations by their impact. For each slot where $V_{\text{slot}}$ = 1, the framework records the $TBS$ of that slot, representing the amount of data that was processed late or potentially dropped. The violated data is accumulated over a sliding window,
$V_{\text{data}} = \textstyle\sum_{i=1}^{n} V_{\text{slot},i} \times \text{TBS}_i$
where n is the window size. This accumulated violated data is then compared against a predefined threshold $V_{\text{threshold}}$: $\text{PHY}_{\text{violation}} = 1$ if $V_{\text{data}} > V_{\text{threshold}}$, and $0$ otherwise. When violated data exceeds the threshold, the violation detection module sends a signal to the configuration manager module to reallocate resources for physical layer. 

Upper layers (MAC, RLC, PDCP, RRC) operate without strict sub-millisecond deadlines. Their processing tasks like MAC scheduling, RLC segmentation/reassembly, and RRC signaling, have relaxed timing constraints on tens to hundreds of (ms) timescales~\cite{varga2024latency}. For these upper layers, CPU utilization serves as a reliable saturation indicator. The framework detects saturation when $\text{CPU}_\ell > \text{CPU}_{\text{threshold}}$ (e.g., 0.95), where $\text{CPU}_\ell$ is per layer utilization. However, this metric fails for PHY layers: moderate CPU utilization (60\%) can still cause TTI violations if workload parallelization is insufficient. This motivates the violated data mechanism for PHY layers.

\subsection{Per-Layer Control Flow Logic}\label{section:adaptive_control}
ALARM receives monitoring metrics, TBS per slot ($\text{TBS}_i$), physical layer processing time ($T_{\text{proc}}$), and per-layer CPU utilization ($\text{CPU}_\ell$), and outputs updated resource configurations $(N_\ell, F_\ell, C_\ell)$ for each layer. The system initializes to a minimal configuration (lowest threads and frequencies), establishing a power-efficient baseline. The control loop operates continuously. At each iteration, the violation detection module computes per layer violation status (e.g., $\text{PHY}_{\text{violation}}$). When violations occur, the configuration manager applies frequency-first scaling: transition from low to high frequency. If already at maximum frequency, increase thread count ($N_\ell$) and invoke core mapping to assign threads based on layer weight ($w_\ell$) and core capabilities ($cap_c$). Proper mapping prevents cache misses and overhead that negate parallelization benefits. When no violations occur, the manager scales down resources.

%========================================================
% Section IV : System Setup, Methodology, Experimental Results
%========================================================

\section{Methodology and Experimental Analysis}
We conduct two experimental studies to evaluate ALARM. The first study assesses the impact of instant per-layer resource allocation on power consumption, the second study evaluates ALARM's dynamic per-layer adaptation. Both studies use the same testbed infrastructure but examine different aspects of the framework under distinct operational modes.

\subsection{Testbed Description} \label{section:testbed}
The gNB is implemented using srsRAN 5G \cite{srsRAN}, deployed on Odroid H4 (8-core Intel Core i3 @ 1.8 GHz) and Raspberry Pi 5 (4-core ARM Cortex-A76 @ 2.4 GHz) single-board computers. USRP B210 is used as RU, a commercial 5G smartphone as UE, and the 5G core network is implemented using Open5GS. To isolate baseband processing effects, high-capacity fronthaul and backhaul links are used, and the UE is placed close to the antennas to reduce the influence of radio-channel impairments. The system uses a 30 kHz subcarrier spacing (TTI = 0.5 ms) with a fixed 64-QAM MCS (index 27) over a single and dual 20 MHz channel. 

\subsection{Quantifying RAN Layers Heterogeneity}

\begin{figure}[ht]
  \centering
  \includegraphics[width=0.95\linewidth]{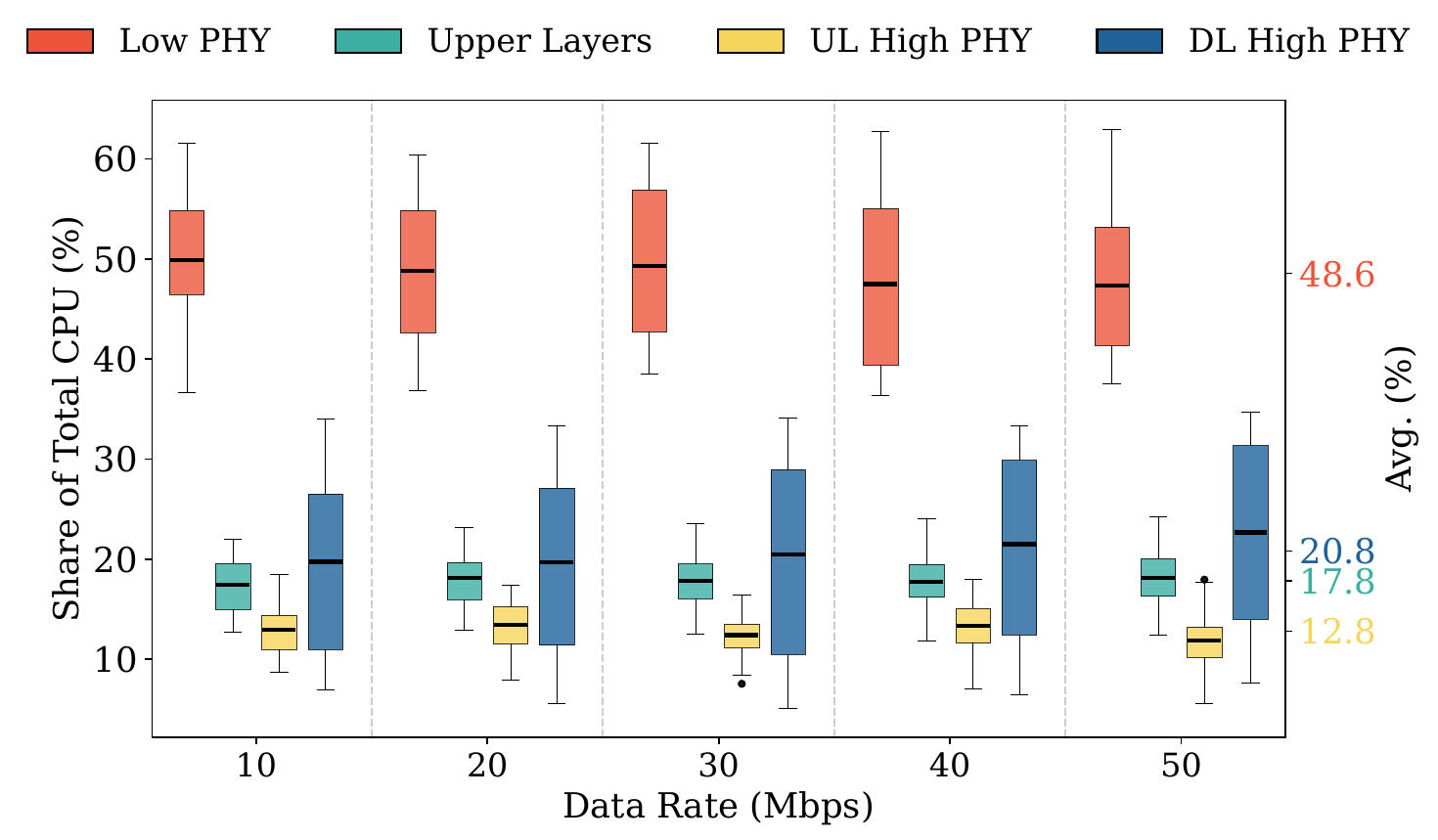}
  \caption{Layer-level CPU utilization vs E2E throughput}
  \label{fig:layers_datarate}
\end{figure}

To quantify the computational heterogeneity across RAN layers, we measured per-thread CPU utilization under varying traffic profiles. We monitored four thread groups: Low-PHY, UL high-PHY, DL high-PHY, Upper Layers. As Fig.~\ref{fig:layers_datarate} demonstrates a constant CPU share for each layer over varying traffic load, physical layer threads collectively consume approximately 82\% of total CPU resources, while upper layer processing accounts for only 18\%. 
\subsection{Study I: Impact of Per-Layer Resource Management} \label{study1}
This study evaluates the power consumption benefits of layer-aware resource management under static traffic conditions. We compare Default and Uniform configurations representing related work~\cite{kalia_towards_nodate, huang_reinforcement_2024, crespo_energy-aware_2025, urumkar_power_2025, urumkar_improving_2023, leonelli_dynamic_2025, lozano_airic_2024} to ALARM configuration:

\textbf{Default:} Baseline without enhancement. Threads managed by OS scheduler, CPU frequency by Linux DVFS, no core affinity. The RAN spawns 18 total threads 11 data plane and 7 auxiliary control threads denoted \textit{ctrl\_th} (radio control, I/O, USB events, gNB control, PUCCH, logging). \textit{ctrl\_th} have minimal CPU impact but increase scheduling complexity.

\textbf{Uniform:} Traditional non layer-aware resource management applied at the gNB entity level. Data plane threads reduced from 11 to 7 and mapped uniformly across cores 1-7 via CPU affinity, treating all RAN processing as a monolithic workload. This configuration represents traditional optimization example that reduces thread count but ignores layer-specific computational heterogeneity.

\textbf{ALARM:} Layer-aware resource management, data plane threads reduced from 11 to 5, allocated per layer as ($N_{\mathrm{L}}$:2, $N_{\mathrm{U}}$:1, $N_{\mathrm{D}}$:1, $N_{\mathrm{Up}}$:1). Each layer thread is mapped to a dedicated core via CPU affinity: ($C_{\mathrm{L}}$:[3,4], $C_{\mathrm{U}}$:6, $C_{\mathrm{D}}$:7, $C_{\mathrm{Up}}$:5). Control threads (ctrl\_th) remain OS-managed on cores 3 and 4. Per-layer frequency scaling was not applied due to hardware limitations, and the default DVFS governor remained active.

\input{figures/power_od_pi}

Fig.~\ref{fig:power_comp} presents the cumulative distribution of power consumption on two constrained platforms. The sharp CDF transitions reflect power shifts between idle and active transmission states. ALARM achieves 29-33\% power reduction versus Default and 11-19\% versus Uniform, demonstrating that traditional vRAN deployments waste nearly one-third of their energy budget on resource over-provisioning, while Uniform reduces thread count without layer awareness, leaving substantial efficiency gains uncaptured. By applying the ALARM concept, power consumption is further minimized while preserving real-time guarantees. To understand the source of these savings, we analyze the system-level results.

\begin{figure}[t]
  \centering
  \includegraphics[width=\linewidth]{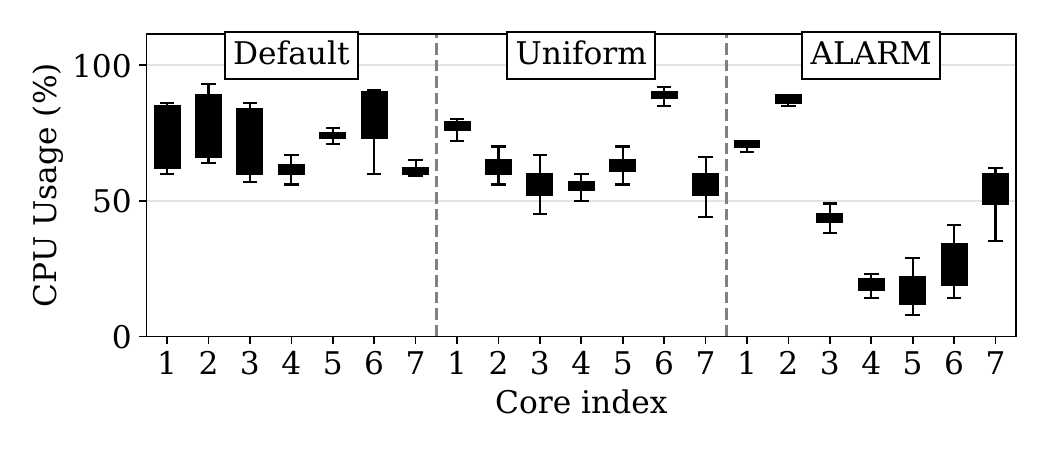}
  \caption{Per-core CPU utilization (\%) comparison}
  \label{fig:cpu_histogram}
\end{figure}

\begin{figure}[ht]
  \centering
  \includegraphics[width=\linewidth]{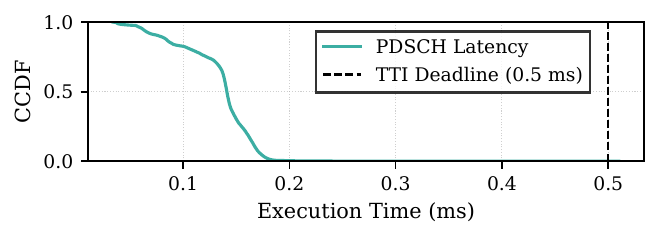}
  \caption{ALARM physical-layer processing time}
  \label{fig:timenoproblem}
  \vspace{-.9em}
\end{figure}

Both platforms show consistent trends, for brevity, we analyze Odroid results. Fig.~\ref{fig:cpu_histogram} presents per-core CPU utilization for each configuration, validating the potential for layer-specific processing reduction. In Default and Uniform scenarios, seven cores exhibit high utilization ranging from 56\% to 93\% in Default and averaging 65.5\% in Uniform, reflecting uncontrolled thread distribution with frequent migrations that spread load evenly across cores. In contrast, ALARM's layer allocation creates functional specialization where each core handles a specific RAN layer, enabling predictable execution patterns. Tab.~\ref{tab:core_layer_gain_dual} demonstrates how ALARM eliminates over-allocation: lightweight layers show 9-78\% CPU reduction, while compute-intensive physical layers retain necessary capacity. The critical question is whether efficiency gains come at the cost of real-time guarantees. Fig.~\ref{fig:timenoproblem} provides a definitive answer. Each point represents a single slot's processing latency under 50 Mbps traffic. Downlink physical layer processing time remains below the 500 $\mu$s TTI deadline for every slot throughout the measurement period, validating that ALARM maintains strict real-time guarantees despite thread minimization.

\input{tables/percore_gain}

\begin{figure}[ht]
  \centering
  \includegraphics[width=\linewidth]{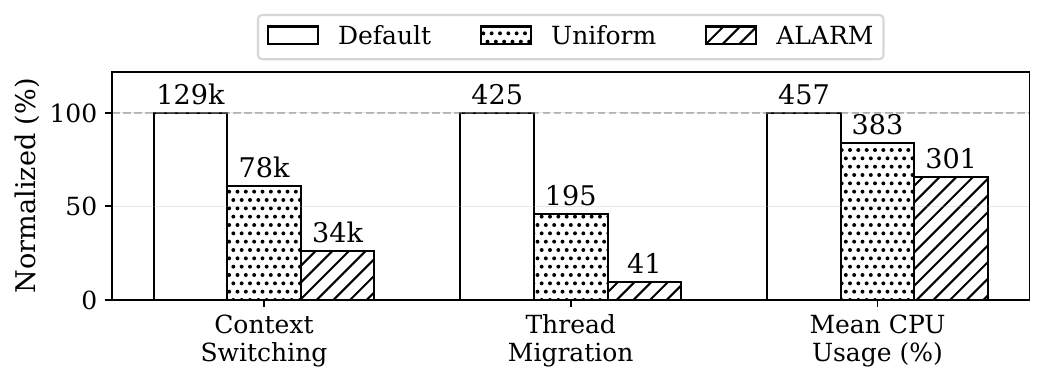}
  \caption{System-level metrics comparison}
  \label{fig:metrics_comparison}
 % \vspace{-.9em}
\end{figure}
Fig.~\ref{fig:metrics_comparison} compares system-level metrics between Default, Uniform, and ALARM, normalized to the Default baseline (100\%). While Uniform successfully reduces total CPU utilization, context switching, and thread migration by reducing the logical executors load, 
ALARM achieves an additional 43.5\% reduction in context switches through per-layer thread minimization and CPU affinity pinning, which eliminates involuntary switching and dramatically reduces OS scheduling activity. Eliminating RAN thread migrations preserves cache locality, thread's data remains resident in CPU's L1/L2 caches rather being evicted and reloaded. This reduces scheduling overhead, manifested in total CPU utilization decreasing from 383\% (Uniform) to 301\% (ALARM). The same workload completes in less time because instructions execute faster with hot caches, and the CPU spends more cycles on actual work.

A crucial observation is that average CPU frequency remains nearly constant at ~1.4 GHz in both scenarios. This proves that power savings do not result only from simply running slower. The same frequency delivers the same throughput with dramatically less overhead. The efficiency gains come entirely from per layer resource allocation, fewer threads, dedicated cores, preserved cache locality, and reduced uncore activity. This is an ALARM's fundamental advantage, it operates at the same performance point while consuming less power, which was validated in the power results shown in Fig. \ref{fig:power_comp}.

\subsection{Study II: Dynamic Per-Layer Adaptation} \label{study2}

Aggressive resource reduction shown in Study I improves power efficiency but risks performance degradation under increasing traffic load, necessitating dynamic adaptation mechanisms. This study compares ALARM's layer-specific adaptation (ALARM Adaptation) against uniform scaling approaches that represent the traditional baseline (Uniform Adaptation)~\cite{kalia_towards_nodate, huang_reinforcement_2024, urumkar_power_2025, leonelli_dynamic_2025}. Default is excluded from this study as it lacks dynamic adaptation capabilities:

\textbf{ALARM Adaptation:} Dynamic response to deadline violations through layer-specific resource adjustment. Manages resources by scaling global frequency to minimum level (1.1 GHz) and adjusting per-layer thread allocation. Thread count $N_\ell$ and core mapping $C_\ell$ are modified in response to the violated layer, while per-layer frequency scaling is unavailable due to hardware limitations. Violation threshold $V_{\text{threshold}}$ is set to 2.5 Mbps, and traffic varies 10-50 Mbps, forcing dynamic adaptation while maintaining real-time guarantees.

\textbf{Uniform Adaptation:} Same configuration as ALARM before adaptation, but scales all layers uniformly when violations occur, rather than targeting only affected layers. 
% Same configuration considered as ALARM before adaptation, but scales all layers uniformly when violations occur as the traffic load increase, rather than targeting only affected layers. 
\input{tables/gain_adaptation}

When violations take place, both adapting approaches resolve violations, however Uniform Adaptation does so at significantly higher power cost. Tab.~\ref{tab:gain_adaptation} shows the power efficiency advantage of ALARM's per-layer adaptation compared to uniform scaling. ALARM achieves at least 9.8\% additional power savings by adapting only to the affected layers rather than all layers simultaneously.

\begin{figure}[ht]
  \centering
  \includegraphics[width=0.95\linewidth]{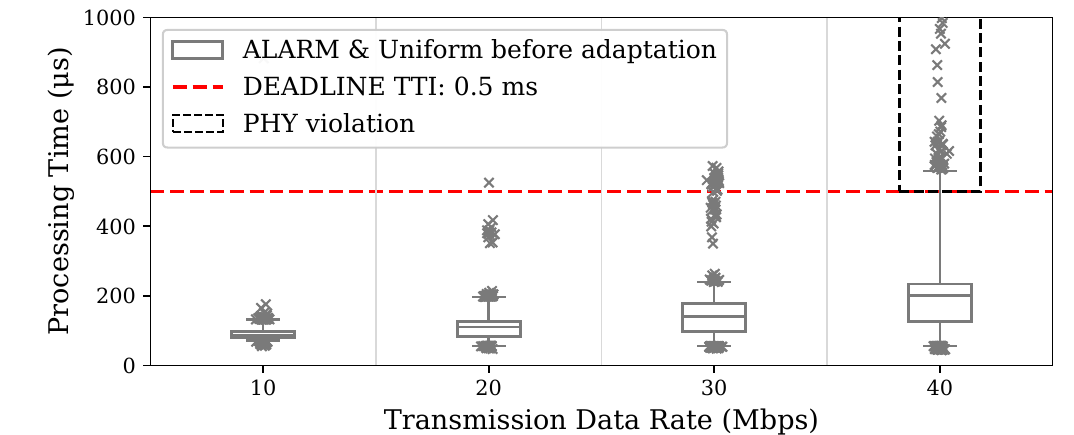}
  \footnotesize
  \caption{Physical layer processing time before adaptation}
  \label{fig:adaptive_a}
\end{figure}

Fig.~\ref{fig:adaptive_a} presents physical layer processing time across traffic scenarios before adaptation takes place. At light to moderate loads (10-30 Mbps, PRB: 14.6-29.1), some slots miss the deadline but violated data remains below the 312.5 KB threshold. These violations represent transient phenomena such as scheduling jitter, interference spikes, or cache misses not capacity shortage. ALARM and Uniform maintain minimal configuration ($N_{\mathrm{L}}$:1, $N_{\mathrm{U}}$:1, $N_{\mathrm{D}}$:1,  $N_{\mathrm{Up}}$:1) across this range, preserving maximum power efficiency.

\begin{figure}[ht]
  \centering
  \includegraphics[width=0.95\linewidth]{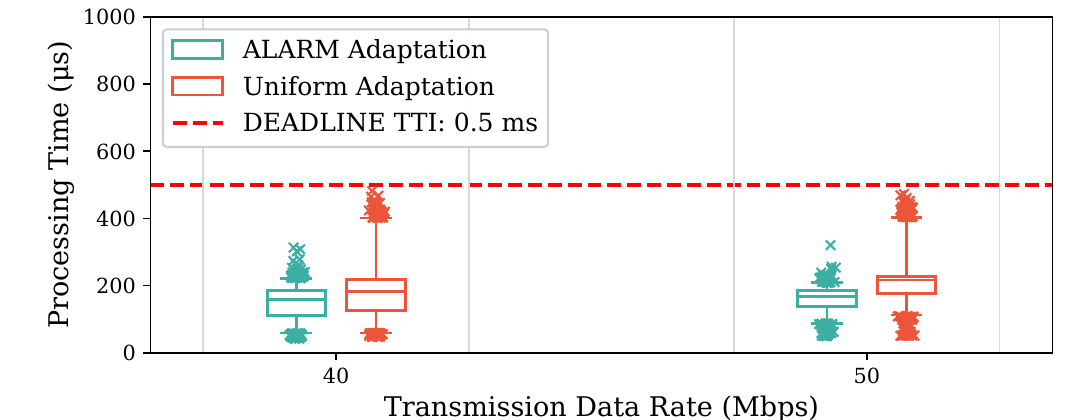}
  \footnotesize
  \caption{Physical layer processing time after adaptation}
  \label{fig:adaptive_b}

\end{figure}

At 40 Mbps without adaptation, the system crosses into capacity shortage. PRB allocation rises to 37.2, violations occur frequently, and violated data reaches 666.3 KB, exceeding the threshold by multiple of 2. The violation detection module computes $V_{\text{data}} = 666.3$~KB, exceeding $V_{\text{threshold}} = 312.5$~KB and thus triggering $\mathrm{PHY}_{\mathrm{violation}}$. This signals genuine capacity exhaustion rather than transient spikes. ALARM responds by addressing only the affected physical layers, increasing $N_{\text{L}}$ and $N_{\text{D}}$ from 1 to 2, then reassigning threads to cores, while Uniform Adaptation increases one thread for all layers to address the problem. The result is immediate as shown in Fig. \ref{fig:adaptive_b} violated data drops to negligible levels and processing times return to the tight distribution seen at lower loads. The adapted configuration ($N_{\mathrm{L}}$:2, $N_{\mathrm{U}}$:1, $N_{\mathrm{D}}$:2,  $N_{\mathrm{Up}}$:1) restores violation-free operation and maintains it through 50 Mbps (PRB: 43.4), validating sustainable reconfiguration. Both scenarios preserve performance, however efficiency is strongly affected by adaptation granularity, Uniform Adaptation allocates resources to unaffected layers, while per-layer adaptation preserves a 9.8\% efficiency advantage by avoiding this unnecessary overhead.

%---small summary of experimental results

\textbf{Discussion:} 
We find that violation severity (violated data volume) provides a more reliable capacity indicator than violation frequency, which primarily reflects transient effects rather than genuine resource shortages. This distinction represents an important consideration for future traffic-aware resource management frameworks.

%========================================================
% Section V : Conclusion
%========================================================
\section{Conclusion}\label{section:conclusion}

This paper presented ALARM, a layer-aware resource management framework for energy-efficient vRAN. By decomposing the gNB into heterogeneous layers with distinct processing demands, the framework enables surgical per-layer resource orchestration rather than monolithic over-provisioning. ALARM provides per-layer control, layer-level violation detection, and dynamic traffic-driven scaling. Using thread allocation to illustrate the approach, experimental validation demonstrates up to 19\% greater power savings than uniform control solutions, with per-layer adaptation achieving 9.8\% additional savings over adaptive uniform scaling approaches. While these magnitudes are hardware-specific and depend on the underlying mechanisms (e.g., frequency scaling, sleep states), our key contribution is demonstrating that layer-aware management provides complementary optimization that enhances, rather than replaces existing power management mechanisms.

Although this work focuses on small-cell deployments, the principles present significant opportunities for larger-scale environments. Cloud-RAN deployments exhibit similar challenges at greater scale, where layer-aware management could enable substantial aggregate power savings. Extending this framework to distributed cloud-based architectures therefore represents an important direction for future research.
\bibliographystyle{IEEEtran}  
\bibliography{reference}
\end{document}

%% file: figures/power_od_pi.tex
\begin{figure}[ht]
    \centering

    \subfloat[Raspberry Pi 5]{%
        \includegraphics[width=0.48\columnwidth]{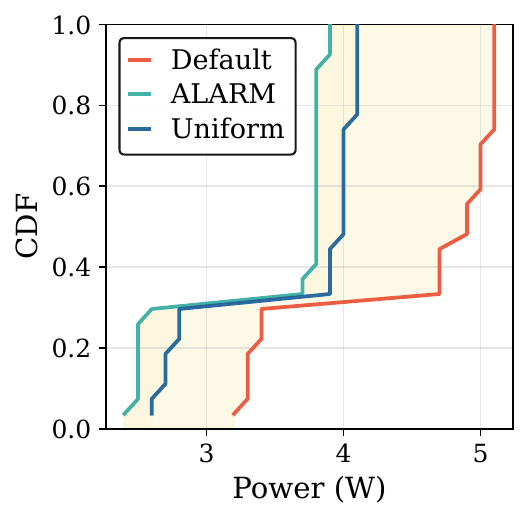}
        \label{fig:pi}
    }
    \hfill
    \subfloat[Odroid]{%
        \includegraphics[width=0.48\columnwidth]{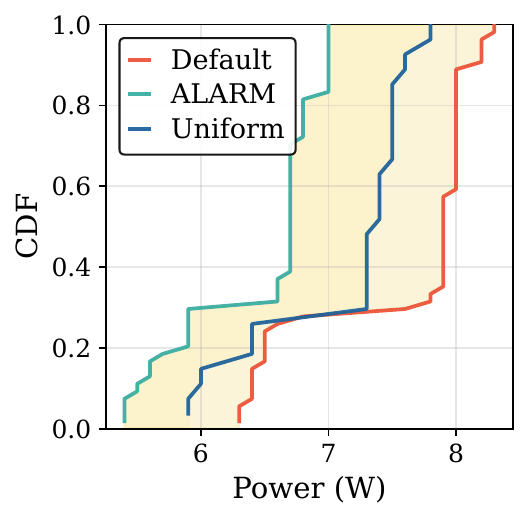}
        \label{fig:odroid}
    }

    \caption{Power consumption comparison of ALARM against Default and Uniform on Raspberry Pi and Odroid platforms}
    \label{fig:power_comp}
    %\vspace{-.9em}
\end{figure}

%% file: tables/percore_gain.tex
\begin{table}[t]
\centering
\footnotesize
\setlength{\tabcolsep}{4pt}

\begin{tabular}{c|ccccccc}
\hline
\textbf{Core} & 1 & 2 & 3 & 4 & 5 & 6 & 7 \\
\textbf{Allocated Layer} & L & L & ctrl\_th & ctrl\_th & Up & U & D \\
\textbf{Default (\%) $\downarrow$ } & 0 & 0 & 38.77 & 69.37 & 77.91 & 67.77 & 17.44 \\
\textbf{Uniform (\%) $\downarrow$ } & 0 & 0 & 24.27 & 65.5 & 74.00 & 70.23 & 8.54 \\
\hline
\end{tabular}

\caption{ALARM's per-core CPU\% reduction.}
\label{tab:core_layer_gain_dual}
\vspace{-.8em}
\end{table}

%% file: tables/gain_adaptation.tex
\begin{table}[ht]

\footnotesize
\centering
\begin{tabular}{cccc}
\hline
\textbf{ALARM (W)} &
\textbf{Uniform Adaptation (W)} &
\textbf{Gain (\%)} \\
\hline
4.9 & 5.4 & 9.8 \\
\hline
\end{tabular}

\caption{ALARM power efficiency vs. uniform adaptation}
\label{tab:gain_adaptation}
%\vspace{-.7em}
\end{table}